\documentclass[]{ceurart}
\usepackage{listings}
\usepackage{color,soul}

\usepackage{amsthm}

\newtheoremstyle{exstyle}
  {5pt} 
  {5pt} 
  {} 
  {} 
  {\bfseries} 
  {.} 
  {.5em} 
  {} 

\theoremstyle{exstyle}
\newtheorem{example}{Example}

\begin{document}

\copyrightyear{2024}
\copyrightclause{Copyright for this paper by its authors.
  Use permitted under Creative Commons License Attribution 4.0
  International (CC BY 4.0).}

\conference{SEBD 2024: 32nd Symposium on Advanced Database System, June 23-26, 2024, Villasimius, Sardinia, Italy}

\title{Machine Learning-Augmented Ontology-Based Data Access for Renewable Energy Data}

\address[1]{Department of Computer Science, University of Milan, Italy}
\address[2]{Department of Information Engineering and Computer Science, University of Trento, Italy}
\author[1]{Marco Calautti}[%
email=marco.calautti@unimi.it,
]
\author[2]{Damiano Duranti}[%
email=damiano.duranti@unitn.it,
]
\author[2]{Paolo Giorgini}[%
email=paolo.giorgini@unitn.it,
]

\begin{abstract}
Managing the growing data from renewable energy production plants for effective decision-making often involves leveraging Ontology-based Data Access (OBDA), a well-established approach that facilitates querying diverse data through a shared vocabulary, presented in the form of an ontology.
Our work addresses one of the common problems in this context, deriving from feeding complex class hierarchies defined by such ontologies from fragmented and imbalanced (w.r.t. class labels) data sources. 
We introduce an innovative framework that enhances existing OBDA systems. This framework incorporates a dynamic class management approach to address hierarchical classification, leveraging machine learning. The primary objectives are to enhance system performance, extract richer insights from underrepresented data, and automate data classification beyond the typical capabilities of basic deductive reasoning at the ontological level. 
We experimentally validate our methodology via real-world, industrial case studies from the renewable energy sector, demonstrating the practical applicability and effectiveness of the proposed solution.
\end{abstract}

\begin{keywords}
Hierarchical Classification \sep 
Ontologies \sep 
Renewable Energy \sep 
Machine Learning
\end{keywords}

\maketitle

\section{Introduction}



The considerable amount of unstructured data produced by heterogeneous facilities across renewable energy production plants presents challenges in effective categorization and analysis. Integrating domain knowledge with data, ontologies can play a central role in facilitating a more effective analysis~\cite{gruber_ontologies}, allowing users to gain richer insights.


A well-established approach for aiding access to raw data by means of ontologies is the so-called Ontology-Based Data Access (OBDA) approach~\cite{ijcai2018p777}, which defines a data access architecture composed of three main levels. At the lowest level, there are the data sources, collecting data regarding the domain of interest, while at the highest level we have instead an ontology, i.e., a formal model of the domain knowledge, that is used to navigate the data using a common vocabulary, closer to domain experts.
In between, 
we have the so-called mappings, which form a specification of how the raw data at the sources should be converted in order to comply with the conceptual model of the ontology. On top of this architecture, we have the users that pose queries over data using only the ontology vocabulary, while the system takes care of converting the request to a request performed over the sources. 
 
In the case of renewable energy production, a data source often coincides with a particular power plant, exporting data describing its internal organization and operations, while the ontology is a formalization of the ISO IEC-81346 standard~\cite{iec81346} which, roughly, indicates that industrial systems (such as a power plant) shall be organized into different subsystems, all of them related by a "part of" relation. In particular, a simpler system (such as a single device) shall be part of a larger, more complex system (e.g., a power production machine), which in turn can form an even more complex system. Moreover, systems at each level shall be classified according to classification schemes dedicated to that level (i.e., a lowest level system could be classified as a temperature sensor, while a higher level system could be an electric motor).
A simplified example of how these components interact, specific to the renewable energy sector, is given below.

\begin{example}\label{ex.obda}
    Assume a power plant exports the following tabular data regarding all its circuit breakers.
    
    \begin{center}
    \begin{tabular}{cccc}
    \toprule
    Device Id & Power Plant & Description \\
    \midrule
    100 &	Power plant 1 & First Engine Circuit Breaker \\
    101 & Power plant 1 & Second Engine Circuit Breaker \\
    \bottomrule
    \end{tabular}
     \end{center}
    
    The IEC-81346 standard dictates that to specify that a subsystem is a circuit breaker one needs to classify this system using the class "QA". Hence, the ontology encoding the IEC-81346 standard contains, among other things, two terms (encoded using the W3C standard OWL ontology language): the term \texttt{iec-81346:PowerPlantComponentQA}, which denotes the class of all circuit breakers, and the term \texttt{iec-81346:ClassifiedAs}, which denotes the relationship between a system and its class.
    	
    Then, a possible mapping converting the given data using the ontology vocabulary is the mapping that executes a query over the above table, searching for all rows mentioning "Circuit Breaker", and for each such row of the form
    	$$(\textbf{id},\textbf{plant}, \ldots \mbox{Circuit Breaker} \ldots),$$
    	it will create a subject-predicate-object statement (triple) of the form
    	$$\langle \textbf{plant/id}\rangle \texttt{ iec-81346:ClassifiedAs } \texttt{iec-81346:PowerPlantComponentQA},$$
    	stating that the system with identifier \textbf{id} in power plant \textbf{plant} is classified as "QA". Such a statement is then stored in a unified database (i.e., a triple store). A mapping, such as the one described above, is usually specified using the W3C standard R2RML language.
    	Then, a user of the OBDA system has access to the statements obtained from the mapping process, and can query the power plants data using the W3C standard SPARQL query language. For example, the following very simple SPARQL query retrieves all circuit breakers:
    \begin{multline*}
            \texttt{SELECT ?x WHERE \{} 
                 \texttt{?x iec-81346:ClassifiedAs iec-81346:PowerPlantComponentQA\}}.
    \end{multline*}
\end{example}

A distinct gap exists in the literature concerning effective ways to map the raw source data to ontological vocabularies. 
Current solutions often rely on manual definitions of such mappings, which is not scalable or efficient for data management and analysis. Furthermore, the dynamic nature of sensors and devices within the domain and their volume present challenges that have not been adequately addressed in the literature.

\smallskip
\noindent
\textbf{Contributions.}
This work introduces a novel approach that implements machine learning models to automate the classification of the source data into the classes of the target ontology, 
providing an automated mean of mapping relevant parts of the source data to the ontology vocabulary. 

Within this framework, a primary challenge lies in effectively managing the extensive hierarchical classification schemes derived from standards such as IEC-81346, which exhibit significant class imbalance. This imbalance arises from the similarity in internal organizational structures among power plants, resulting in a scarcity of examples for less common classes of systems. To address this issue, we introduce a novel dynamic class management approach, allowing for optimized model performance by leveraging the inherent hierarchical structure of the underlying data.
We demonstrate the real-world applicability of our approach through a case study developed with our industrial partner, Enyr\footnote{\url{https://www.enyr.eu/}}, a leading company in data modeling in the renewable energy sector. We point out that, while we focus on a case study about renewable energy production, the methodology and framework proposed are general enough to be applied across various sectors.



\section{Related work}\label{sec:related-work}

%

OBDA has been recognized as a standard and sound approach for data integration for its semantic clarity, flexibility in integrating new data sources \cite{lenzerini2011ontology}, and efficiency exploited by specialized algorithms and solutions~\cite{calvanese2017ontology}.
OBDA is central when data heterogeneity and the dynamism of sources introduce significant challenges. However, its application in specific domains, such as renewable energy, introduces certain limitations. Particularly, these limitations become apparent when the source data requires additional preprocessing before aligning with the ontological schema. A notable example is the presence of unstructured data (e.g., text) at the sources, requiring an extra layer of preprocessing to seamlessly integrate into the ontological framework. Indeed, it is extremely common that data exported by production plants do not follow any well-established standard, while at the same time, they describe objects organized in complex hierarchies.
To the best of our knowledge, no OBDA architecture exists in the literature that can handle the above limitations, especially for the renewable energy sector.
A promising way of dealing with the above issues is to employ machine learning techniques.

The convergence of ontologies and machine learning is mainly evident in three distinct research areas: ontology embeddings for machine learning, ontology learning, and ontology population.
\begin{itemize}
    \item Ontologies are often encoded via languages such as OWL, the Web Ontology Language defined by the W3 Consortium, and define formal descriptions of entities and their relationships. To feed ontologies to machine learning methods, they first need to be transformed into mathematical structures, a process known as ontology embedding. Techniques like On2Vec \cite{on2vec} and OWL2Vec* \cite{owl2vec} facilitate this, leveraging graph-based methods and language models with ontology graphs. 
    Many of those approaches focus on embedding-based relation prediction or combining language models with random walks on the ontology graph. 
    \item Ontology learning is the process of automatically generating ontologies from data using machine learning techniques. These applications typically involve Natural Language Processing (NLP) techniques for extracting relevant terms and concepts from text. Subsequent phases involve machine learning to model knowledge structures and relationships. Some applications highlight the potential of automated relation extraction towards a more automated approach to learning conceptual relations from text~\cite{girju_discovering,pantel_ontologizing,petit_discovering}. 
    Studies also showed how automatically extracted relationships from unknown data could compete and outperform approaches based on semantic taxonomies and manually-encoded sets of relations \cite{asium, unsupervised,learning_patterns}. 
    \item Ontology population is the process of creating instances for an ontology from a data source. Traditional population processes often employ rule-based techniques, which may require expert validation both of the rules and the extracted elements to solve consistency problems \cite{Amaral2013RulebasedNE, Cimiano2005TowardsLO, FinkelsteinLandau1999ExtractingSR, AlaniAuto2003, Makki2008OntologyPV}. These solutions are resource-intensive, involving considerable manual effort to establish and update, particularly when adapting to changes in data structures. Alternative approaches integrate Natural Language Processing (NLP) to better exploit language constraints \cite{witte-etal-2010-flexible, ruiz-martinez, labidi2017}. Some solutions introduce machine learning to enhance the ontology population process, and are mainly focused towards relation extraction from text \cite{suchanek-etal-2006-leila, zhou-etal-2005-exploring, zeng-etal-2014-relation}. However, these approaches use linguistic assumptions to identify lexical structures or patterns, which are then used to extract relational data. While these methods mark a significant advancement, they face limitations in dealing with short and unstructured text, lacking any formal structure.
 
\end{itemize}
\smallskip
\noindent 
The first two approaches above have gained remarkable results in combining machine learning with ontologies, while the third, which is closer to the focus of our work, has seen limited applications of machine learning techniques to ontologies, and OBDA systems in general.

The goal of our work 
is to propose a methodology to populate an existing ontology, i.e., map the source data of an OBDA system to the classes of the underlying ontology, by automatically classifying data at the source using machine learning techniques. In doing so, our approach effectively exploits the class hierarchies of the ontology.

\section{The Renewable Energy Data Management Challenge}\label{sec:enyr}

The case study we consider in this work, which will serve as a specific instance of our framework, is provided by our industrial partner, Enyr, a company specializing in data modeling for the renewable energy sector, particularly modeling data produced by renewable energy production plants. Their experience has shown that these plants, characterized by their diverse range of devices, components, and sensors, generate a large volume of data, and thus, the challenge lies in integrating and standardizing these data across different sources and systems. They have found that, although international data organization standards, such as the IEC-81346 specifications~\cite{iec81346}, have already been available since years, the data governance strategies adopted in this sector have been fragmented, with individual plants operating in relative isolation, and thus relying on internal policies (if any) on how to store power plants-related data.

This approach is no longer efficient due to the recent expansion of the renewable energy sector.
Hence, translating and mapping existing data into such standard models is of paramount importance, and presents different challenges, such as volume and variety of source data, as well as lack of data structure (i.e., use of free text) at the sources.







The aforementioned challenges underscore the need for new methods to enhance a traditional OBDA system by integrating data-driven techniques like machine learning. This integration is crucial for the efficient mapping of data from power plants to an IEC-81346-based ontological schema.



\section{MLA-OBDA Architecture}\label{sec:methodology}

\begin{figure}[t]
	\centering
	\includegraphics[width=0.85\textwidth]{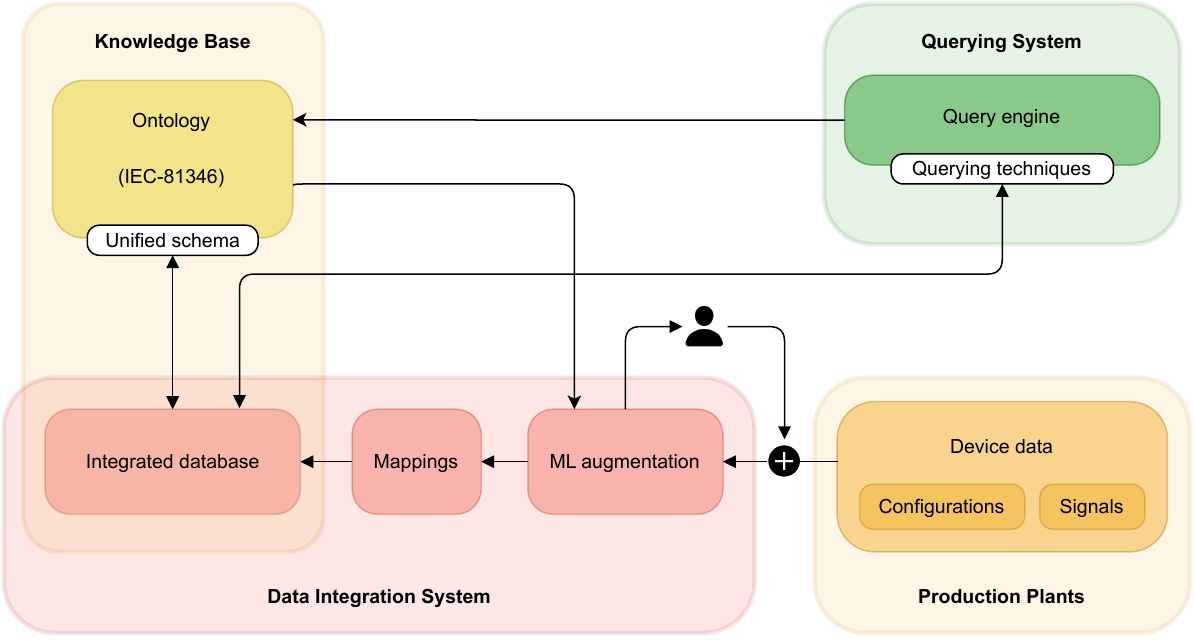}
	\caption{The ML-OBDA architecture} \label{fig1}
\end{figure}

Our proposal involves augmenting a traditional OBDA system with a machine learning-based component integrated between the data sources and the data mappings. This augmentation aims to bolster the system's capability, ensuring a more comprehensive feed of data into the ontological schema.
The proposed architecture, called \emph{machine learning augmented}, or MLA-OBDA for short, is shown in Figure \ref{fig1}. Below, we outline the key components of the architecture, with the initial three being somewhat standard in an OBDA system.

\smallskip
\noindent
\textbf{Production Plants.} 
Starting at its base, production plants and facilities generate raw data, representing the internal system organization. Following its generation, this raw data is directed towards the “Data Integration” system. At this stage, the goal is to map the raw data to a format adhering to the domain specification, which will be stored in an integrated database.

\smallskip
\noindent
\textbf{Knowledge Base.}
Centrally within this architecture is the knowledge base, which integrates the data obtained from the data sources with the ontological constructs that comply with established standards such as IEC-81346. 
The flow from the “Data Integration” system to the “Knowledge Base” in Figure \ref{fig1} represents the continuous alignment between stored data and the ontological constructs.

\smallskip
\noindent
\textbf{Querying System.}
A crucial component in any OBDA system is the querying system, which is closely integrated with the knowledge base, facilitating users in extracting meaningful insights from the data now encoded using the ontology vocabulary. This is a standard component of any OBDA system that acts as an interface to query across different data sources and is able to rewrite queries posed over the ontology vocabulary to queries over the integrated database in a transparent way, e.g., see~\cite{calvanese2017ontology}.

\smallskip
\noindent
\textbf{Data Integration System.}
Fully exploiting the potential of a unified ontological schema necessitates effectively integrating current data into our outlined framework. This part of the architecture now presents a novel component, i.e., the "ML Augmentation" component, whose goal is to preprocess the incoming raw data from plants in order to enrich it, before the mappings can convert the data to one that follows the ontology specification. The novel ML-based component is composed of the following elements:

\begin{itemize}
    \item\textbf{Machine Learning Model}: This is responsible for the automatic classification, according to the ontology class hierarchies, of the different subsystems of the power plants, which are encoded using raw, unstructured data;
    \item\textbf{Mappings}: This describes the process of reconciling the classified data towards the unified ontological schema;
    \item\textbf{Integrated database}: This stores the data encoded according to the ontological schema.
\end{itemize}

The decision to integrate a machine learning layer into our methodology was driven by the need for a more sophisticated mechanism to process and classify existing data, ensuring it aligns seamlessly with the ontology; we note that a separate component of the whole architecture is the domain expert, which can interact with the ML augmentation component in order to provide feedback needed to improve the classification models, such as rectifying some wrong classifications. In the next section, we discuss how we implemented the ML augmentation component.

\section{Machine Learning Augmentation}\label{sec:augmentation}


This section describes how different machine learning models are used to classify data about entities of the system, using the classification hierarchies coming from the ontology. The key problem is to classify a (sub)system given a dataset coming from the data sources (i.e., power plants), where each system is represented by means of a free text description. The classification has to follow the classification hierarchies specified by the IEC-81346 ontology. We provide the following details on how the IEC-81346 standard (and thus the corresponding ontology) specifies how systems are organized and classified and how the data produced by the power plants is shaped. Then,  Section~\ref{sec:classification} presents the overall classification approach.

\smallskip
\noindent
\textbf{Subsystem classification.}
%
According to the IEC-81346 standard, a system inside a power plant can be classified into one of 3 main classes, representing, roughly, the "scope" of the system, i.e., whether it is a simple, low-level system, such as a device, or a more complex system. Such classes are called \emph{breakdown levels (BL)}, which represent different degrees of granularity in the functional breakdown of the system; BL0, BL1, BL2 denote the three breakdown levels, where BL0 denote systems of the highest scope (i.e., larger systems), while BL2 denote systems at the lowest scope (e.g., a device).
Each system of a certain BL is further classified using a class coming from a class hierarchy specific to that BL. Each such hierarchy is a mean to classify a system at different levels of details. For example, a system at BL2 could be classified as an "electric lamp" (finest level of granularity in the classification hierarchy), or more generally as a "light object", or even more generally as an "emitting object". Such classification hierarchies inside a certain BL represent an additional challenge, as they differ in structure, depending on the BL. Systems at BL0 are classified using a "flat" classification scheme, i.e., there is only one granularity level of classification, while systems at BL1 and BL2 use a deeper classification hierarchy up to three levels.

\smallskip
\noindent
\textbf{The Power plants Data.}
Systems of a power plant are attached a human-written textual description, which will be used as the input for the classification task. These texts exhibit significant variability in length and structure, suggesting diverse patterns in data. Each dataset shows a significant class imbalance. The observed imbalance is primarily due to limited data for certain classes rather than an overrepresentation of others. Current data distributions represent real-world conditions, as noted by our industrial partner, suggesting that the imbalance is an inherent aspect of the problem, and an artificially balanced dataset might perform poorly when faced with real-world, imbalanced data. Several strategies were considered to address data imbalance. Resampling methods, such as oversampling to artificially inflate minority class samples or undersampling to decrease majority class samples, are standard practices. However, with text data, oversampling may lead to overfitting, while undersampling could result in the loss of significant information \cite{yap_oversampling,mohammed_oversampling}. We then propose a different approach to class imbalance that we are going to discuss in the next subsection.

\subsection{Dynamic Class Management}\label{sec:classification}

At the highest breakdown level (BL0), the task is a classical text classification problem, where the target is to classify data using classes from a "flat" pool of classes. The application of the selected models aims to maximize the classification performance, and each class is considered equally important.

At the second (BL1) and third level (BL2), a significant factor that adds to the complexity of the problem is the distribution of data across the classes. The strong data imbalance is a direct consequence of these levels' internal classification hierarchies. Each subclass in the hierarchy adds additional information to the parent one. It is crucial to understand if and when such internal structure should be ignored or exploited.

The potential approaches to face such a classification challenge are either to \emph{1)} "flatten" the classification hierarchy to a pool of classes, where each class is equally important, or to \emph{2)} actually take into account the classification hierarchy.

Our proposal is to \emph{combine} flat classification principles with a dynamic class management approach, as several tests within the domain showed how the classification accuracy of a model on a specific class grows with a growing number of samples due to higher generalization capabilities. From the nature of the data, we can derive that lower-level classes, carrying the maximum detail, but with few samples, may be merged into their parent class to enhance the model’s pattern recognition and predictive capabilities. 

\smallskip
\noindent
\textbf{Methodology.}
In what follows, fix a breakdown level $B \in \{\mbox{BL1},\mbox{BL2}\}$. The goal is to understand whether a certain class $C$ in the hierarchy of $B$ is "represented enough" by the data so that a classification model can be safely trained on such data, or instead, we should merge the samples of class $C$ to the ones of class $C'$, where $C'$ is the parent of $C$ in the hierarchy. We achieve the above as follows:
\begin{enumerate}
	\item Choose a threshold \textit{v} on the minimum number of samples for each class $C$ in the dataset that is deemed "enough" for training; we will clarify what we mean by "enough" later.
	\item \textbf{while} there is a class $C$ with less than $v$ samples in the dataset \textbf{do}
	\begin{enumerate}
		\item modify the class of the samples with class $C$ in the dataset to the class $C'$, which is the parent of $C$ in the hierarchy.
	\end{enumerate}
	\item Train a classifier on the obtained dataset.
\end{enumerate}

\begin{figure}[t]
	\centering
	\includegraphics[width=0.55\textwidth]{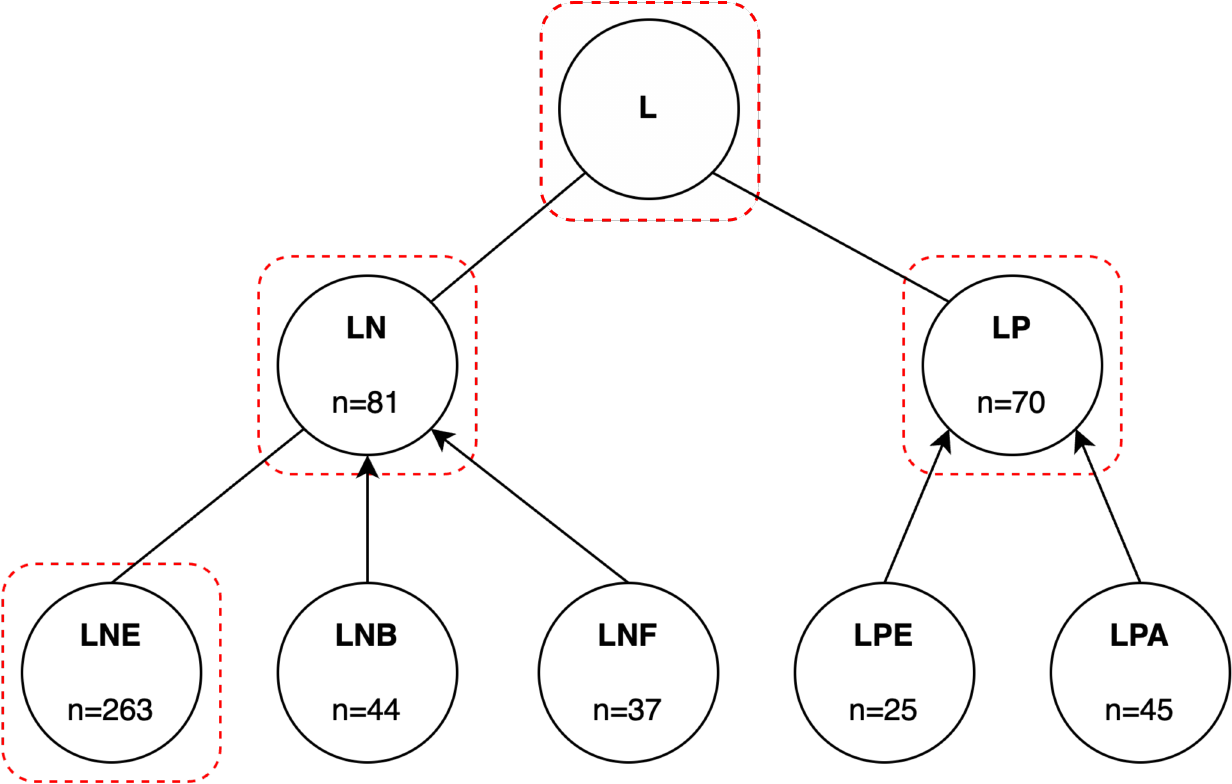}
	\caption{Example application of the proposed approach, where \textit{v} = 50, and the number $n$ of samples of each class is reported for each class. The classes that are kept for classification are circled in red.} \label{fig2}
\end{figure}

The goal of the above methodology is to merge samples of a less represented class $C$ (i.e., with less than \textit{v} samples) to samples of the parent class $C'$. Moreover, if after the change, even the class $C'$ is still not well-represented, then its samples are further merged to the parent of $C'$, and so on. Being able to choose the value of \textit{v} allows to strike a balance between higher average performance of the trained model and the granularity of classification. Figure \ref{fig2} shows an example of the approach, where the tree represents an extremely simplified version of the classification hierarchy for BL2, where each node is a class of the hierarchy; for clarity, the class is annotated with the number $n$ of samples of that class in the dataset. With a threshold \textit{v} = 50, the classes inside a red circle are the ones that remain in the dataset after applying the discussed approach, while all samples of the other classes are merged to the samples of more general classes up the tree.\footnote{In the picture, each class is represented by its IEC-81346 code; the number of letters in the code denotes the class level. For example, the top class \textbf{L} denotes "Steam, water, condensate systems", while its subclass \textbf{LN} denotes "Impounding systems for hydroelectric power plant".}

What remains to clarify is how to choose the value of \textit{v}, i.e., what do we mean by "enough for training"?
%
We determine the value of the threshold \textit{v} as the minimum one required to achieve the maximum value \textit{t} of a desired average performance metric of the trained model. The most valuable and employed metric in a multiclass context, where the target is increasing the performance of the underrepresented classes, is the Macro F1-score~\cite{grandini_metrics}.


\begin{table*}
  \caption{Model performance for classifying BL0 systems. Here, 'P' stands for Precision, 'R' for Recall, and 'F1' for the F1 Score.}
  \label{tab:bl0}
  \begin{tabular}{lccccccccc}
    \toprule
    & \multicolumn{3}{c}{\textbf{BERT}} & \multicolumn{3}{c}{\textbf{Naive Bayes}} & \multicolumn{3}{c}{\textbf{Random Forest}} \\
    \cmidrule(r){2-4} \cmidrule(r){5-7} \cmidrule(r){8-10}
    & \textbf{P} & \textbf{R} & \textbf{F1} & \textbf{P} & \textbf{R} & \textbf{F1} & \textbf{P} & \textbf{R} & \textbf{F1} \\
    \midrule
    \textbf{Weighted} & 0.99 & 0.99 & 0.99 & 0.96 & 0.96 & 0.96 & 0.98 & 0.98 & 0.98 \\
    \midrule
    \textbf{Macro} & 0.99 & 0.99 & 0.99 & 0.90 & 0.95 & 0.92 & 0.97 & 0.96 & 0.97 \\
    \midrule
    \textbf{Accuracy} & \multicolumn{3}{c}{0.99} & \multicolumn{3}{c}{0.96} & \multicolumn{3}{c}{0.98} \\
    \bottomrule
  \end{tabular}
\end{table*}

\section{Evaluation}\label{sec:evaluation}

This section evaluates the approach over real-world, manually validated data, provided by our industrial partner.\footnote{Due to the confidentiality of the data, we were not allowed to make it public.} The evaluation is performed focusing on traditional metrics like Precision, Recall, and F1-score. Hence, our ground truth comprises 50,125 text entries, each describing a device from some renewable energy power plant. The power plants are owned by customers of our industrial partner, and each text has been manually labeled with the appropriate device class by domain experts from our industrial partner.

The results were computed on a validation set of 10,025 of the 50,125 entries, corresponding to 20\% of the available dataset, while the models were trained on the remaining 40,100 samples; each entry (record) describes in text form a (sub)system of a power plant. Specifically, the ‘Description’ column, the input for classification, exhibited significant variability, composed of 42,851 unique descriptions of variable length, with an average of 6.71 words per entry. 

Fields ‘BL0’, ‘BL1’, and ‘BL2’, containing the class of the system at the corresponding breakdown level, were composed of one, three, and two letters, respectively, meaning the second and third present a class at lower levels of the corresponding hierarchy, and thus are suitable to apply our dynamic class management approach. All breakdown levels displayed significant class imbalance. For instance, systems at BL0 have five unique classes, BL1 has 137, and BL2 has 55. The unbalanced distribution is evidenced with certain classes, such as \textbf{M} (denoting systems for conversion of energy) in BL0, which constituted approximately 72.18\% of entries.

Since each entry of our dataset consists of free text, generating mappings manually for the dataset is unfeasible. This, together with the availability of manually validated entries, further justifies the use of traditional machine learning metrics, such as precision and recall.

\smallskip
\noindent
\textbf{Data preprocessing.}
We perform two main steps of data preprocessing. We first perform a data cleaning pass, whose goal is to remove noise from input data, i.e., we uniform lowercase and uppercase text and remove frequent and non-informative patterns, including excessive spaces and symbols (the latter is essential in our context, due to the prevalence of abbreviations in the dataset). Lastly, we filter out data irrelevant to the classification task, such as those representing internal component enumerations within plants (e.g., "Engine \textbf{1}" vs "Engine \textbf{2}"). Then, we apply standard text tokenization
to decompose raw text input into smaller, manageable "chunks".
After this preprocessing phase, we move to the model selection.

\begin{figure}[t]
    \centering
    \begin{minipage}{0.5\textwidth}
        \centering
        \includegraphics[width=\linewidth]{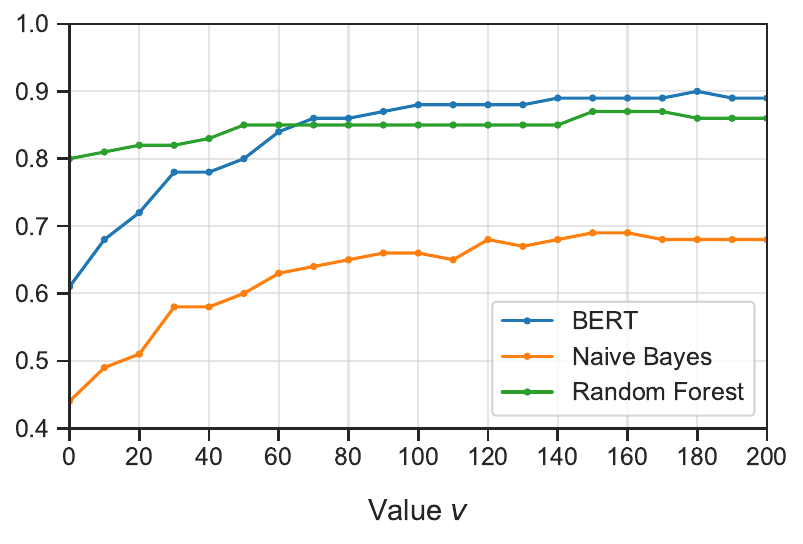} 
    \end{minipage}\hfill
    \begin{minipage}{0.5\textwidth}
        \centering
        \includegraphics[width=\linewidth]{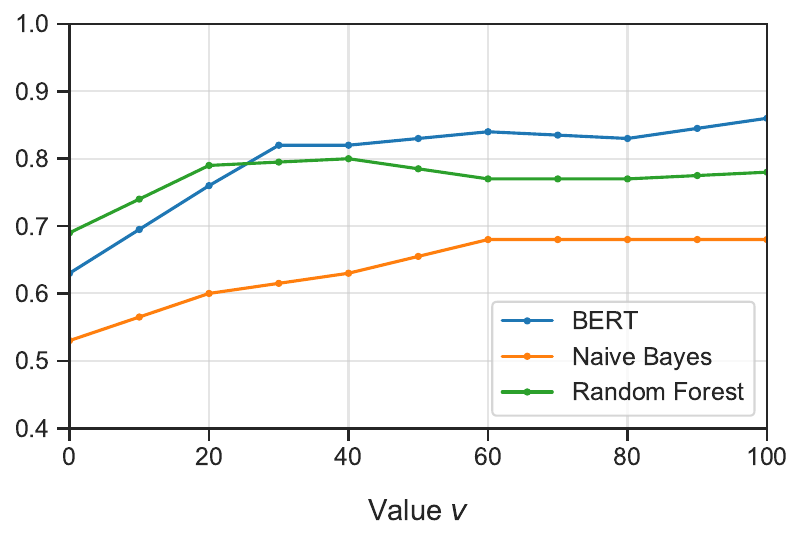} 
    \end{minipage}
    \caption{Evolution of Macro-F1 score with different \textit{v} values on BL1 (left) and BL2 (right).}
    \label{fig3}
\end{figure}

\smallskip
\noindent
\textbf{Model selection.}
For the models, we considered BERT~\cite{bert}, Naive Bayes, and Random Forests. These models vary significantly from one another and therefore were selected to more accurately assessing the generalizability of the proposed solution.

BERT is a state-of-the-art transformer-based model known for its capabilities in processing natural language data with a bidirectional approach, which allows for a deep understanding of the context within text data. Naive Bayes is a probabilistic classifier and provides a straightforward approach to classification. Random Forests are known for their high accuracy and ability to handle large datasets with numerous input variables. 

The hyperparameters of each model were tuned through an extensive combination trial employing a grid search. The optimal parameter configurations for each model proved to be the same through all the breakdown levels due to the similarity of the tasks. 

The values for the BERT base uncased model were selected among the parameters most used in the literature, where typical values for learning rate range between 3 $\times$ 10\textsuperscript{-5} and 3 $\times$ 10\textsuperscript{-4}. Recommended values for batch sizes vary from 8 to 64, depending both on the problem and the available hardware, and the optimal number of epochs is dependent on the task, still being typically few and between 1 and 5 \cite{komatsuzaki_one_epoch,sun_finetune_bert}. Combining a learning rate of 4 $\times$ 10\textsuperscript{-5}, a batch size of 32, and 5 epochs, without weight decay, using the Cross-Entropy loss function, resulted in the most efficient combination for this specific task; the model employed its tokenizer.

The optimal configuration of the Naive Bayes model employed an ’alpha’ value of 0.01, while the Random Forest was configured with 600 trees. For both models, data was preprocessed through CountVectorizer, which converts text into a matrix of token counts.

\begin{table*}
  \caption{Consolidated Classification Reports on BL1 Validation Set. 'P' represents Precision, 'R' represents Recall, and 'F1' represents the F1 Score.}
  \label{tab:bl1}
  \begin{tabular}{llccccccccc}
    \toprule
    & & \multicolumn{3}{c}{\textbf{BERT}} & \multicolumn{3}{c}{\textbf{Naive Bayes}} & \multicolumn{3}{c}{\textbf{Random Forest}} \\
    \cmidrule(r){3-5} \cmidrule(r){6-8} \cmidrule(r){9-11}
    & & \textbf{P} & \textbf{R} & \textbf{F1} & \textbf{P} & \textbf{R} & \textbf{F1} & \textbf{P} & \textbf{R} & \textbf{F1} \\
    \midrule
    \multirow{2}{*}{\textbf{Weighted}} & \textbf{Flat} & 0.95 & 0.94 & 0.94 & 0.81 & 0.81 & 0.79 & \textbf{0.91} & \textbf{0.91} & 0.90 \\
    & \textbf{Dynamic} & 0.95 & 0.94 & \textbf{0.95} & \textbf{0.83} & \textbf{0.82} & \textbf{0.81} & 0.90 & 0.90 & 0.90 \\
    \midrule
    \multirow{2}{*}{\textbf{Macro}} & \textbf{Flat} & 0.61 & 0.63 & 0.61 & 0.55 & 0.41 & 0.44 & 0.86 & 0.82 & 0.80 \\
    & \textbf{Dynamic} & \textbf{0.86} & \textbf{0.90} & \textbf{0.88} & \textbf{0.82} & \textbf{0.63} & \textbf{0.68} & \textbf{0.89} & \textbf{0.84} & \textbf{0.85} \\
    \bottomrule
  \end{tabular}
\end{table*}

\smallskip
\noindent
\textbf{Classification Performance.}
In the following evaluation, the terms 'Weighted' and 'Macro' refer to the different methods of averaging the performance metrics (Precision, Recall, F1-Score). The 'Weighted' rows in the tables present averages that are weighted by the number of instances in each class. On the other hand, the 'Macro' rows show unweighted averages, which is an approach for better understanding model performance across all classes. Additionally, the 'Flat' and 'Dynamic' labels distinguish the two classification strategies. The 'Flat' approach refers to conventional classification where all classes present in the dataset are used for training, while 'Dynamic' refers to the approach presented, that dynamically adjusts the classes attached to the samples, before training is performed.

As already discussed, for systems at BL0 there is no need for our dynamic class management approach, since the classes for this level are not organized in a hierarchy. Hence, in Table \ref{tab:bl0} we simply report the performance of the three models trained and validated on the untouched dataset. It is evident that employing a transformer-based model such as BERT generally provides higher performance.

We now consider BL1 and BL2. Here, the goal is to improve average model performance across the largest set of classes rather than on majority classes only. The results show improved average performance after applying the dynamic approach described in the previous section. We have chosen the value of \textit{v} by analyzing the evolution of the macro-F1 score value while varying the value \textit{v}, and chose the value of \textit{v} that led to a macro-F1 score that stabilizes above a certain threshold \textit{t}. 
Figure \ref{fig3} shows this analysis. The minimum value \textit{v} of 0 means no classes were merged. However, we point out that this does not imply that every class was considered, as a minimum of 10 samples per class was still needed to perform training and testing, and classes with less than 10 samples were discarded from the dataset, regardless of the value of \textit{v}.

For the BERT-based approach applied on BL1, the macro-F1 score was observed to stabilize once it reached 88\%, with a minimum \textit{v} of 100 samples per class. This value reduced the classes to 63. In particular, 6 of the lowest level classes were merged to their direct parent, while other 11 of the lowest level classes had to be merged at their ancestor two levels up the hierarchy. The same analysis was performed for the other models employed and resulted in selecting two values \textit{v} of 120 and 50 samples for the Naive Bayes and Random Forest models, respectively. This approach enabled training the models on some less-represented higher-level classes, which on their own did not reach the minimum amount of 10 elements. The performance of the models was then measured on the validation set, and the results are shown in Table \ref{tab:bl1}.

We applied our dynamic class management approach also to BL2; Figure \ref{fig3} shows an optimal threshold \textit{v} for BERT of 30 elements per class, 20 for Random Forest, and 60 for Naive Bayes. Applying these values resulted in the performance gain shown in Table~\ref{tab:bl2}.


\begin{table*}
  \caption{Consolidated Classification Reports on BL2 Validation Set. In this table, 'P' denotes Precision, 'R' denotes Recall, and 'F1' denotes the F1 Score.}
  \label{tab:bl2}
  \begin{tabular}{llccccccccc}
    \toprule
    & & \multicolumn{3}{c}{\textbf{BERT}} & \multicolumn{3}{c}{\textbf{Naive Bayes}} & \multicolumn{3}{c}{\textbf{Random Forest}} \\
    \cmidrule(r){3-5} \cmidrule(r){6-8} \cmidrule(r){9-11}
    & & \textbf{P} & \textbf{R} & \textbf{F1} & \textbf{P} & \textbf{R} & \textbf{F1} & \textbf{P} & \textbf{R} & \textbf{F1} \\
    \midrule
    \multirow{2}{*}{\textbf{Weighted}} & \textbf{Flat} & 0.99 & 0.99 & 0.99 & 0.90 & 0.89 & 0.88 & 0.83 & 0.83 & 0.83 \\
    & \textbf{Dynamic} & 0.99 & 0.99 & 0.99 & 0.90 & 0.89 & 0.88 & 0.83 & 0.83 & 0.83 \\
    \midrule
    \multirow{2}{*}{\textbf{Macro}} & \textbf{Flat} & 0.64 & 0.65 & 0.63 & 0.56 & 0.52 & 0.53 & 0.71 & 0.68 & 0.69 \\
    & \textbf{Dynamic} & \textbf{0.82} & \textbf{0.82} & \textbf{0.82} & \textbf{0.76} & \textbf{0.66} & \textbf{0.68} & \textbf{0.82} & \textbf{0.78} & \textbf{0.79} \\
    \bottomrule
  \end{tabular}
\end{table*}

Despite a slight loss of information, the increased balancing of performance achieved through the hierarchical classification approach underscores the benefits of employing a tailored threshold solution, particularly in class-imbalanced datasets coupled with complex hierarchical structures. Consolidating minority classes has proven effective in enhancing overall performance metrics, particularly in transformer-based models, highlighting the adaptability and robustness of the approach. On the other hand, the Random Forest model's performance was shown to be fairly stable and not influenced by the approach. This can be explained by considering two key aspects. Firstly, Random Forests present an ensemble approach of combining multiple decision trees, which may lead to mitigating problems deriving from class imbalance. Moreover, this approach shows an ability to capture complex relationships, which means that it can perform well even without data restructuring, even with its default hyperparameter settings. However, the highest performance is still achieved by combining the BERT-based model with our dynamic class management approach, providing for BL1 and BL2 a 40\% and 28\% improvement, respectively,  across all metrics.

%
%

\section{Future Steps}\label{sec:conclusion}


We presented a novel integration of machine learning techniques with OBDA systems, via a dynamic class management approach that enhances the population of ontologies.

Since the proposed approach loosely relies on the specifics of OBDA, as it only requires data to be organized in \emph{finite} hierarchies, we would like to explore the application of our approach to other areas of knowledge representation, such as feeding facts to logic programs with finite models~\cite{CalauttiGST15,CalauttiGT13,CalauttiGMT15}. Moreover, considering that source data, in OBDA settings, could be often inconsistent w.r.t.\ domain-specific integrity constraints (e.g., see~\cite{CaCP19,CalauttiGMT22,CalauttiCGMTZ21,CaCP21}), it would be interesting to study extensions of the proposed approach that take into account such inconsistencies, in order to improve accuracy.

\begin{acknowledgments}
\vspace*{-2mm}
This work was funded by the European Union - Next Generation EU under the MUR PRIN-PNRR grant P2022KHTX7 “DISTORT”, and by the Autonomous Province of Trento, under the Provincial Law no. 6, December 13th 1999, grant C79J22002130001 "n-Model".
\end{acknowledgments}

\bibliography{bibliography}

\end{document}